\documentstyle{cupconf}

\ifoldfss
\else
  \ifnfssone
    \newmathalphabet{\mathit}
      \addtoversion{normal}{\mathit}{cmr}{m}{it}
      \addtoversion{bold}{\mathit}{cmr}{bx}{it}
    \newmathalphabet{\mathcal}
      \addtoversion{normal}{\mathcal}{cmsy}{m}{n}
    \else
    \ifnfsstwo
    \fi
  \fi
\fi

\def\hexnumber#1{\ifcase#1 0\or1\or2\or3\or4\or5\or6\or7\or8\or9\or
 A\or B\or C\or D\or E\or F\fi }

\makeatletter
\ifx\CUP@mtlplain@loaded\undefined
\else
\fi
\makeatother
 \makeatletter
 \ifx\CUP@mtlplain@loaded\undefined
   \font\tenbmi=cmmib10 at 10pt
   \font\sevenbmi=cmmib10 at 7pt
   \font\fivebmi=cmmib10 at 5pt

   \newfam\bmifam
   \textfont\bmifam=\tenbmi
   \scriptfont\bmifam=\sevenbmi
   \scriptscriptfont\bmifam=\fivebmi
   
 \fi
 \makeatother
\ifnfsstwo

\fi
\ifnfssone

\fi
\ifoldfss

\fi

\mathchardef\varLambda="0103

\makeatletter
\ifx\CUP@mtlplain@loaded\undefined
\else
\fi
\makeatother
\makeatletter
\ifx\CUP@mtlplain@loaded\undefined
  \font\tenbms=cmbsy10
  \font\sevenbms=cmbsy10 at 7pt
  \font\fivebms=cmbsy10 at 5pt
  \newfam\bmsfam
  \textfont\bmsfam=\tenbms
  \scriptfont\bmsfam=\sevenbms
  \scriptscriptfont\bmsfam=\fivebms

  \edef\bsy@{\hexnumber\bmsfam}
  \mathchardef\bnabla="0\bsy@72
\fi
\makeatother
\title[Post-AGB Stars as Standard Candles]{Post-AGB Stars as Standard 
Extragalactic Candles}

\author[H. E. Bond]{Howard E. Bond}

\affiliation{Space Telescope Science Institute, 3700 San Martin Dr., 
Baltimore, MD 21218, USA}

\begin{document}
\ifnfssone
\else
  \ifnfsstwo
  \else
    \ifoldfss
      \let\mathcal\cal
      \let\mathrm\rm
      \let\mathsf\sf
    \fi
  \fi
\fi

\maketitle



\input{epsf.sty}{\relax}


\def\eps@scaling{.95}
\def\epsscale#1{\gdef\eps@scaling{#1}}

\def\plotone#1{\centering \leavevmode
    \epsfxsize=\eps@scaling\columnwidth \epsfbox{#1}}

\def\plottwo#1#2{\centering \leavevmode
    \epsfxsize=.45\columnwidth \epsfbox{#1} \hfil
    \epsfxsize=.45\columnwidth \epsfbox{#2}}

\def\plotfiddle#1#2#3#4#5#6#7{\centering \leavevmode
    \vbox to#2{\rule{0pt}{#2}}
    \includegraphics{#1}}


\begin{abstract}

Stars evolving off the asymptotic giant branch and passing through spectral
types F and A are excellent candidates for a new extragalactic standard
candle. These post-AGB (PAGB) stars are the visually brightest members of
Population~II systems.  They should have a narrow luminosity function, bounded
from above by the shorter transition times of more massive and more luminous
remnants, and from below by the core mass corresponding to the
lowest-mass stars that are leaving the main sequence. 

Moreover, PAGB A-F supergiants are easily recognized because of their enormous
Balmer jumps, and should lie both in ellipticals and the halos of spirals.  I
describe a photometric system that combines the Gunn $u$ filter (lying below
the Balmer jump) with the standard Johnson-Kron-Cousins {\it BVI\/}
bandpasses, and report a successful search for PAGB stars in the halo of M31
using this {\it uBVI\/} system. 

The zero-point calibration will come from PAGB A and F stars in Galactic
globular clusters. Four are presently known, and have a mean $M_V=-3.4$ with a
scatter of only 0.2~mag. Two are in the same cluster, NGC~5986, and their $V$
magnitudes differ by only 0.09~mag, strongly suggesting a narrow luminosity
function.  Adopting this $M_V$ and calculating the M31 distance from its halo
PAGB stars, we exactly reproduce the accepted value. 

Future plans include a {\it uBVI\/} survey of all Milky Way globular clusters
for PAGB stars in order to strengthen the zero-point calibration, and a survey
of Local Group galaxies to check the calibration. Ultimately we believe we can
reach the Virgo Cluster through a distance ladder with only three rungs:
subdwarf parallaxes, Milky Way globular clusters, and then directly to Virgo
(with {\it HST}). 

\end{abstract}

\firstsection 

\def\BD{{BD~$+39^\circ4926$}}
\def\HST{{\it HST}}
\def\popi{Population~I}
\def\popii{Population~II}
\def\Schonberner{Sch\"onberner}
\def\squig{\sim\!\!}
\def\subsun{_{\odot}}
\def\Teff{T_{\rm eff}}
\def\wCen{$\omega$~Cen}

\section{Introduction} 

As is well known, the zero point of the extragalactic distance scale depends
heavily on just one method---Cepheid variable stars. With the continuing
conflict between stellar ages and the age of the Universe implied by the
Hubble Constant, it is worth checking the zero point with as many independent
methods as possible. 

I will argue in this paper that a class of ``\popii'' stars, evolving off the
asymptotic giant branch (AGB), may constitute a superb new class of standard
stellar candles.  I will explain why these ``post-AGB'' (PAGB) stars are
expected to have a narrow luminosity function as they evolve through spectral
types F and A, point out that they should lie in systems that do not possess
Cepheids or complicated interstellar extinction (elliptical galaxies, and
halos of spirals), and show how they can be calibrated within the Milky Way.
Thus these PAGB stars should become an important new {\it primary distance
indicator}, which, I will argue, may prove to be of comparable utility to the
Population~I Cepheids. 

My collaborators in this work are Laura K. Fullton, Abhijit Saha, and Karen
Schaefer (all at STScI), and Rex Saffer (Villanova University). 

\section{``\popii'' Standard Candles} 

Here we use the term ``\popii\ candles'' to refer to distance indicators that
arise from old stellar populations. Among the best-known \popii\ candles are
the RR~Lyrae variables.  Their zero points can be calibrated in the Galaxy
through various means (e.g., statistical parallaxes, the Baade-Wesselink
method, and globular clusters), and thus they can be considered to be primary
distance indicators. They are useful for measuring distances within the Local
Group (see the reviews by Jacoby et al.\ 1992 and van den Bergh 1992, and a
remarkable series of papers by Saha, Hoessel, and collaborators---Saha et al.\
1992 and references therein).  However, at $M_V \approx +0.6$, RR~Lyrae stars
are too faint for ground-based detection outside the Local Group. Even \HST\/
can go only a small distance beyond the Local Group at this brightness level. 

There are other, brighter \popii\ stellar distance indicators.  These include
(1)~long-period variables (cf.~Feast 1996 and references therein);
(2)~non-variable AGB stars (e.g., the carbon-star luminosity function---see
Pritchet et~al.\ 1987; Brewer, Richer, \& Crabtree 1995); and (3)~the
luminosity of the red-giant tip (Lee, Freedman, \& Madore 1993; Madore \&
Freedman 1995; Madore, this workshop), which is now being used to find
distances well outside the Local Group (e.g., \HST\/ has detected the
red-giant tip in NGC~5128---Soria et~al.\ 1996).  The method of
surface-brightness fluctuations (SBFs; see Tonry \& Schneider 1988; Jacoby
et~al.\ 1992; Tonry, this workshop) is also based on the luminosities of red
giants in old populations. The planetary-nebula luminosity function (PNLF;
Jacoby et~al.\ 1992; Jacoby, this workshop) technique is based on the
descendants of AGB stars of low to intermediate mass. 

Many authors, including several at this workshop, have listed the requirements
for a standard candle. A good outline is given by Aaronson \& Mould (1986),
who list the following criteria (to which I have added parenthetical remarks
that apply specifically to {\it stellar\/} candles): 

\begin{itemize}
\item Small scatter (e.g., a small range in $M_V$)
\item Available over wide distance range (i.e., high luminosity)
\item Minimal corrections (e.g., reddening unimportant; weak metallicity
dependence) 
\item Objective measurables (e.g., stellar magnitudes)
\item Physical basis (e.g., a basis in theoretical evolutionary tracks)
\end{itemize}

\noindent to which one can add as additional desiderata: 

\begin{itemize}
\item Easily recognizable objects (i.e., requiring a minimum of scarce
telescope time)
\item Calibratable within our Galaxy (i.e., {\it primary\/} distance 
indicators)
\end{itemize}

Although the various \popii\ candles mentioned above have proven extremely
useful, they nevertheless do not necessarily satisfy {\it all\/} of the
Aaronson-Mould criteria. For example, RR~Lyrae stars, as noted, are not of
high luminosity, and they (as well as LPVs) require a long time-series of
telescope time for their detection. The absolute magnitudes of red giants, AGB
stars, and planetary nebulae cover a wide range, so one must go deep enough,
and detect enough objects, to recognize changes in their luminosity functions
(i.e., the red-giant or AGB tips, or the turnover near the bright end of the
PNLF)\null. The PNLF does not have a primary zero point established within the
Milky Way (due to our very poor knowledge of the distances of individual
Galactic PNe), and the SBF method was likewise not initially based on a
Galactic calibration (although Ajhar \& Tonry 1994 have recently tested the
method with observations of Galactic globular clusters). 

By contrast, I will argue that PAGB stars, as they pass through spectral types
F and A, satisfy {\it all\/} of the standard-candle criteria itemized above. 
They are thus a particularly ``clean'' candle, which will allow us to step out
to the Virgo Cluster (with \HST\/) on the basis of a remarkably small number
of ``rungs'' in the distance ladder. 

\section{PAGB A- and F-Type Supergiants as Standard Candles} 

As a low- to intermediate-mass star nears the tip of the AGB, the mass of its
hydrogen-rich envelope decreases due to nuclear burning from below and
stellar-wind mass loss from above.  When the envelope mass reaches
$\squig10^{-2}M\subsun$, the star leaves the AGB and evolves rapidly across
the HR diagram toward higher effective temperatures. On a time scale of
a few times $\squig10^4$~yr or less, $\Teff$ reaches $\squig30,000$~K and the
surrounding AGB wind is ionized by stellar UV radiation, producing a planetary
nebula (PN)\null. 

Our proposed new extragalactic candles are the transition objects of spectral
types A and F, located between the AGB and the realm of PN nuclei. They lie in
the approximate temperature range $5,000\,{\rm K} < \Teff < 10,000\, \rm
K$\null. 

The top panel of Fig.~1 shows typical theoretical PAGB evolutionary tracks in
the HR diagram, for several remnant masses. (The plotted tracks are taken from
interpolations by L.~Stanghellini between the well-known Sch\"onberner and
Paczynski PAGB tracks---see Stanghellini \& Renzini 1993.) The PAGB stars
evolve across the HR diagram at constant luminosity, with higher $\log
L/L\subsun$ corresponding to higher remnant mass.  (The transition of the PAGB
star across the HR diagram is so rapid that there is essentially no evolution
in the luminosity of its core; as first discovered by Paczynski 1971, the
luminosity is simply a function of the core mass.) Also shown schematically is
the Cepheid instability strip.  The dashed rectangle shows the approximate
location of the A- and F-type PAGB stars with which we are concerned here;
they are blueward of the instability strip.  The vertical dashed line is
located at 30,000~K and marks the transition from PAGB star to
planetary-nebula nucleus (PNN)\null. 

\begin{figure}
\plotfiddle{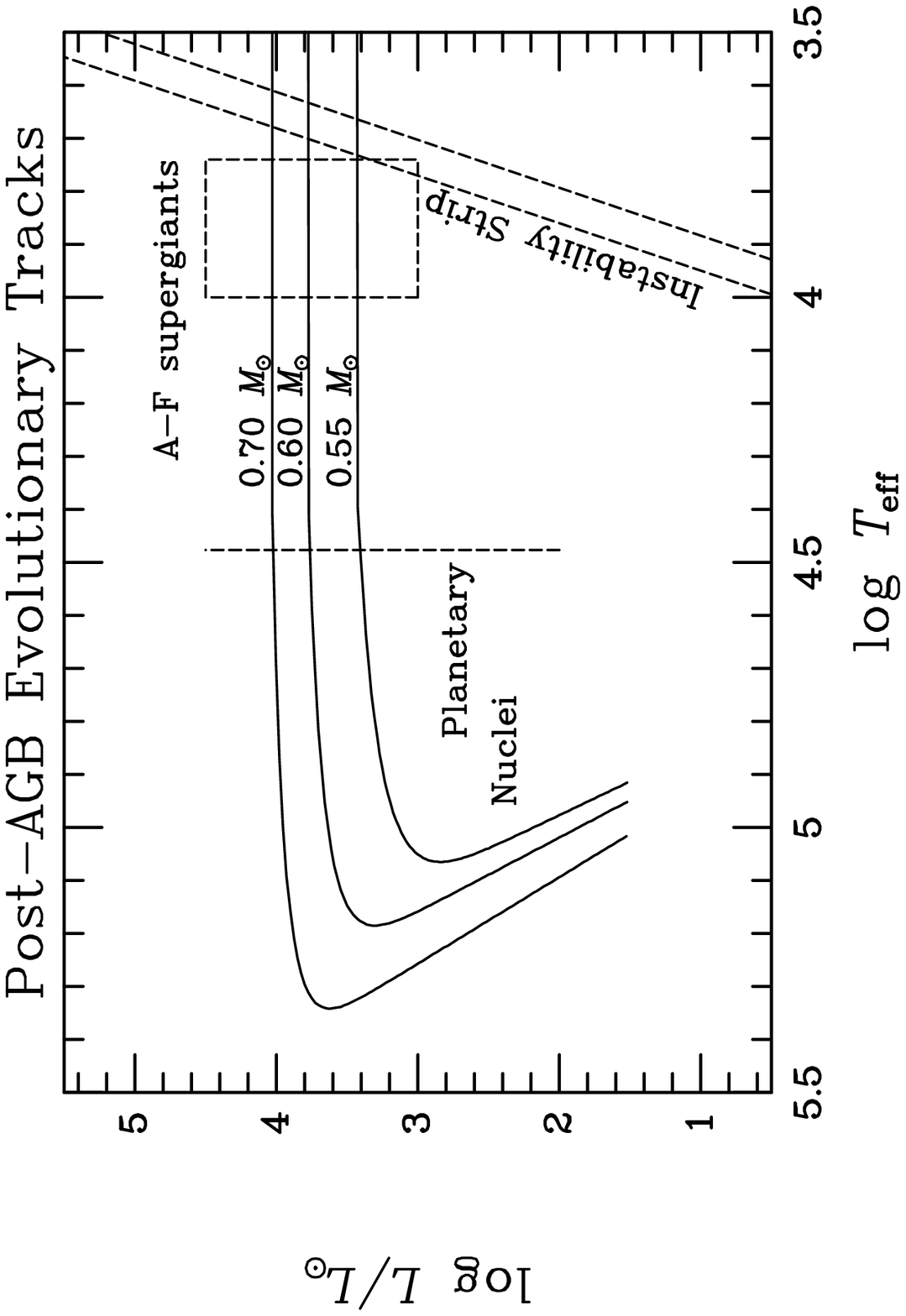}{2.7in}{-90.}{50}{50}{-205}{260}
\plotfiddle{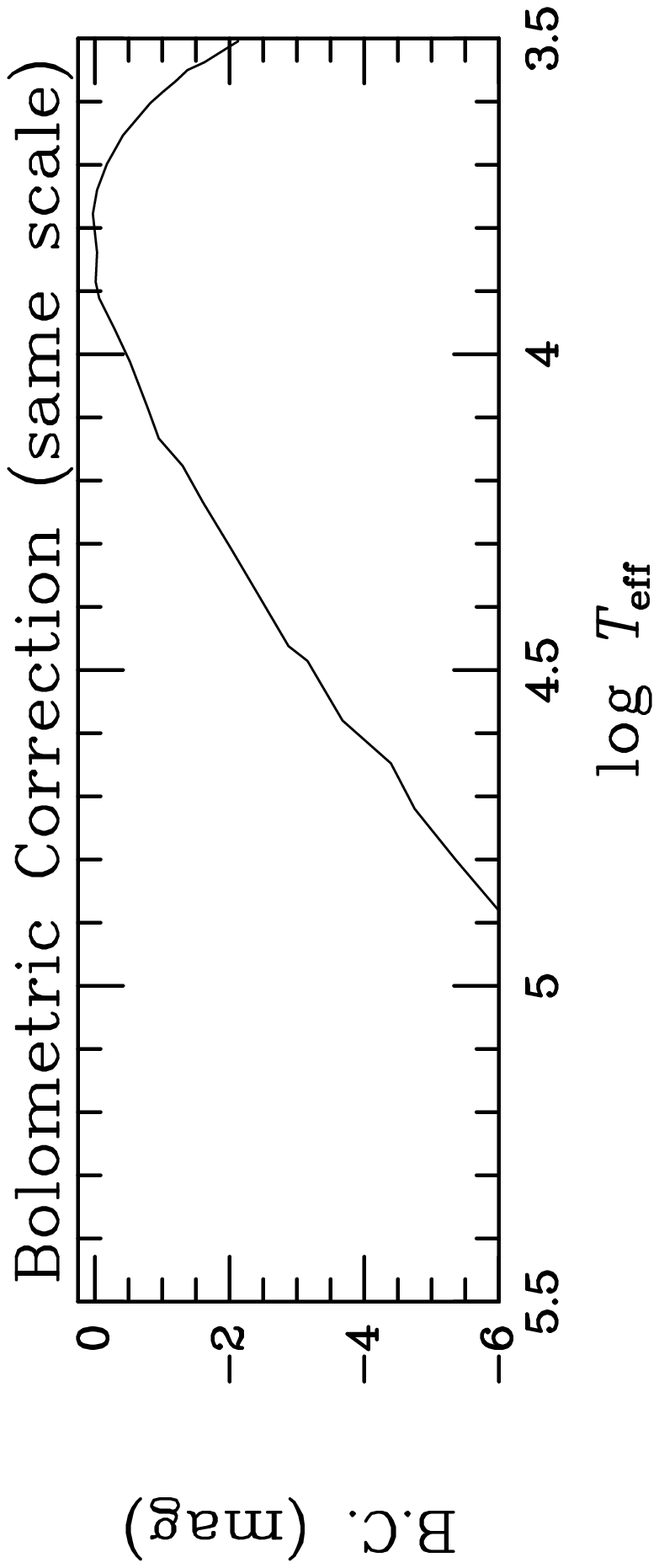}{1.3in}{-90.}{50}{50}{-205}{210}
\caption{({\bf Top}). Schematic HR diagram showing PAGB tracks for three
remnant masses, the pulsational instability strip, the location where
ionization of the circumstellar wind creates a planetary nebula (vertical
dashed line), and the location of the A- and F-type PAGB supergiants discussed
here (dashed rectangle). 
({\bf Bottom}). The bolometric correction, plotted to the same scale as the
top figure, but in magnitude units. This curve shows the shape the PAGB tracks
will have in a plot of $V$ magnitude vs.\ $\Teff$\null.  PAGB stars are at
their greatest visual brightness as they pass through spectral types F and A,
and are fainter both near the AGB and when they reach high temperatures. {\it
PAGB A and F stars are the visually brightest members of \popii.}} 
\end{figure}

The bottom panel of Fig.~1 plots the bolometric correction (from Kurucz model
atmospheres) against effective temperature.  Since the PAGB tracks are
horizontal in $\log L/L\subsun$, this plot shows the shape of the tracks in
the $V$ magnitude.  Because the bolometric correction is smallest for PAGB
stars of types late F and early A, they are {\it the visually brightest
members of \popii}\null. The hotter PAGB stars fade rapidly as $\Teff$
increases above $\squig10,000$~K\null. (Hot PAGB stars have been discovered in
several globular clusters---see de~Boer 1987 for a summary of ground-based
searches, and Dixon et~al.\ 1994 and references therein for space-based work.
However, the strong dependence of the B.C. upon $\Teff$, combined with the
weak temperature dependence of $B-V$, makes it unlikely that the PAGB stars
above 10,000~K will be useful candles.) 

Moreover, we expect the luminosity function (LF) for PAGB A and F stars to be
quite narrow, for the following reasons. (1)~A very sharp lower cutoff should
exist, corresponding to the lowest-mass stars in the stellar population which
are currently leaving the main sequence.  For an old population, such as in
the halo of a spiral galaxy, this lower cutoff will correspond to PAGB stars
of approximately $0.55 M\subsun$, which are the descendants of main-sequence
stars of $\squig0.8M\subsun${}. (2)~The upper cutoff of the PAGB LF is set by
the shorter transition times for more massive remnants. Much more rapid
evolution at higher masses and luminosities was a general property of the
earlier calculations of PAGB evolution (e.g., Paczynski 1971; \Schonberner\
1983). Actually, however, the transition times are extremely dependent upon
the adopted mass-loss laws and are thus rather uncertain; see the detailed
discussions by Bl\"ocker \& Sch\'onberner (1990), Vassiliadis \& Wood (1994),
and Bl\"ocker (1995). Nevertheless, in the Vassiliadis \& Wood (1994)
calculations, evolutionary timescales at $\Teff\approx10,000$~K range from
10,000~yr at $0.569M\subsun$ down to 100~yr at $0.754M\subsun$ (these are the
times taken to go from $\log\Teff=4.0$ to 4.5). Bl\"ocker's recent isochrones
(1995, Fig.~12) also indicate more rapid evolution at higher luminosities.
(3)~Quite apart from considerations of transition times across the HR diagram,
the more massive and brighter remnants will be rarer in a population with a
range of stellar ages, due to the smaller incidence of more massive
progenitors, resulting simply from the initial mass function. 

In a population containing only old stars, the theoretical tracks tell us that
{\it the PAGB LF should shrink almost to a delta function}.  If true, this
would mean in the context of measuring extragalactic distances that we would
only have to {\it detect\/} the PAGB stars; it would not be necessary to go
much deeper than the detection, as is necessary in methods such as the
red-giant tip or PNLF{}. 

\section{Field PAGB A and F Type Supergiants} 

Significant numbers of field PAGB A and F stars are now known in the solar
neighborhood, e.g., from objective-prism surveys for high-latitude
supergiants, or from optical identifications of {\it IRAS\/} sources. 
Examples of those with $\Teff$ and $\log g$ determined from high-resolution
model-atmosphere analyses are listed in Table~1.  Also included is the bright
A-type supergiant, ROA~24 = HD~116745, in the globular cluster \wCen.  As the
table shows, the \wCen\ star's parameters are indistinguishable from those of
the field PAGB candidates. 

\begin{table}
\begin{center}
\begin{tabular}{lccl}
{Star} & {$\Teff$} & {$\log g$} & {Reference}\\
\noalign{\smallskip}
HD 187885     & 8000 & 1.0 & Van Winckel et al.~1996a\\
HD 133656     & 8000 & 1.2\rlap{5} & Van Winckel et al.~1996b\\ 
HD 44179      & 7500 & 0.8 & Waelkens et al.~1992\\
\BD\          & 7500 & 1.0 & Kodaira et al.\ 1970\\
HR 4049       & 7500 & 1.0 & Lambert et al.~1988\\
HR 6144       & 7200 & 0.5 & Luck, Bond \& Lambert 1990\\
IRAS 07134+1005 & 7000 & 0.1 & Klochkova 1995\\
$\omega$ Cen 24 & 6950 & 1.2 & Gonzalez \& Wallerstein 1992\\
HD 161796     & 6600 & 0.0 & Luck, Bond \& Lambert 1990\\
IRAS 18095+2704 & 6600 & 1.0 & Klochkova 1995\\
HR 7671       & 6600 & 1.4 & Luck, Bond \& Lambert 1990\\
89 Her        & 6550 & 0.6 & Luck, Bond \& Lambert 1990\\
HD 56126      & 6500 & 0.5 & Parthasarathy et al.~1992\\
HD 46703      & 6000 & 0.4 & Luck \& Bond 1984\\
HD 52961      & 6000 & 0.5 & Waelkens et al.~1991\\
HR 4912       & 6000 & 0.6 & Luck, Lambert \& Bond 1983\\
RU Cen        & 6000 & 1.1 & Luck \& Bond 1989\\
IRAS 22272+5435 & 5600 & 0.5 & Zacs, Klochkova \& Panchuk 1995\\
\end{tabular}
\end{center}
\smallskip
\caption{Field PAGB A-F Supergiants with Atmospheric Analyses}
\end{table}

Fig.~2 shows the positions of these stars in the $\log g$ vs.\ $\Teff$
diagram, along with the PAGB evolutionary tracks transformed to these
coordinates. This comparison shows that the atmospheric parameters of the
field stars (and the \wCen\ star) are fully consistent with their lying on
standard PAGB tracks. 

\begin{figure}
\plotfiddle{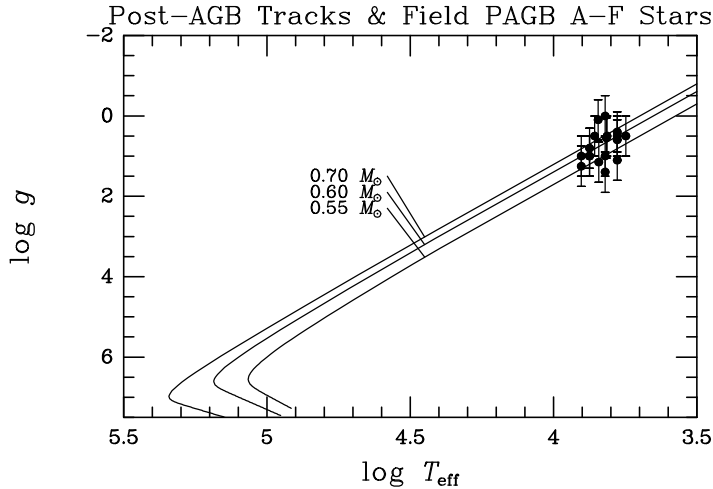}{2.6in}{-90.}{50}{50}{-205}{250}
\caption{Positions of the field PAGB stars from Table~1 in the
gravity-temperature plane. Also shown are the evolutionary tracks from the top
panel of Fig.~1, transformed to the $\log g$,$\Teff$ plane. The positions of
the field stars are consistent with them lying on PAGB tracks.} 
\end{figure}

Unfortunately, the spectroscopic $\log g$ values of the field PAGB stars are
not accurate enough to calibrate their absolute luminosities, even if their
masses were known {\it a priori}.  Therefore it is necessary to base an 
empirical calibration on PAGB stars in globular clusters. 

\section{Calibration of PAGB A-F Supergiants via Globular Clusters}

A few PAGB A- and F-type stars are known in globular clusters, and can serve
for a preliminary calibration of their absolute magnitudes. The well-known
10th-magnitude A supergiant in \wCen, already mentioned above, is the only
star in any globular cluster with its own HD number; its proper motion and
radial velocity leave little doubt that it is a cluster member (e.g., Sargent
1965). Almost 20 years ago, I discovered two PAGB candidates in the
little-studied globular cluster NGC~5986 in the course of a photographic
slitless-spectroscopic survey (Bond 1977); they were readily recognizable
because of their enormous Balmer jumps (see below), and I subsequently
obtained radial velocities confirming their membership. 

Harris, Nemec, \& Hesser (1983) have compiled a listing of stars in 29
globular clusters that are known or suspected to lie above the horizontal
branch and to the left of the red-giant branch.  Their Fig.~3 presents a
composite HR diagram for these stars, and provides an additional candidate for
a non-variable luminous F star, ZNG~5 in M19 (NGC~6273); but to the best of
my knowledge there are no proper motions or radial velocities confirming its
membership. 

Table~2 lists some details for these four objects, including, in the final
column, the absolute visual magnitudes calculated from the cluster distances
and foreground reddening tabulated by Webbink (1985). 

Table~2 dramatically confirms our expectation that the PAGB stars in old
populations will have a very narrow LF: the scatter in $M_V$ among the four
stars is only 0.2~mag---considerably less than the range among Cepheids at a
given pulsation period. Moreover, the mean $M_V$ of $-3.4$ is in remarkable
agreement with the luminosity of the lowest-mass ($0.546M\subsun$) PAGB track
of Sch\"onberner (1983). 

\begin{table}
\begin{center}
\begin{minipage}{5 in}
\begin{tabular}{lcccccccc}
Cluster & NGC & Star & $V$ & $B-V$ & Ref.\footnote{Photometry 
 references: (1) Cannon \& Stobie 1973; (2) Bond unpub.;
 (3) Harris et al.~1976. Reddenings and distances from Webbink 1985.}
 & $E(B-V)$ & Distance & $M_v$\\
 & & & & & & & (kpc) & \\
\noalign{\smallskip}
\wCen & 5139 & ROA 24 & 10.80 & 0.36 & (1) &
  0.11 & \phantom{1}5.2 & $-3.1$\\
\noalign{\smallskip}
\hfil$\dots$      & 5986 & Bond 1 & 12.48 & 0.72 & (2) &
  0.25 & 10.5 & $-3.4$\\
             & $''$ & Bond 2 & 12.39 & 0.51 & $''$             &
  $''$      & $''$                    & $-3.5$\\
\noalign{\smallskip}
M19          & 6273 & ZNG 5  & 12.89 & 0.58 & (3) &
  0.38 & 10.6 & $-3.4$\\
\noalign{\smallskip}
             &      &        &       &      &                    &
       &                    & \llap{Mean\ }$-3.4$\\
             &      &        &       &      &                    &
       &                    & \llap{$\sigma$\ }$\phantom{-}0.2$\\
\end{tabular}
\end{minipage}
\end{center}
\medskip
\caption{Absolute Magnitudes of PAGB A-F Supergiants in Galactic Globular
Clusters} 
\end{table}

The case of NGC~5986 is particularly instructive. This hitherto obscure
globular cluster seems destined to fill a role comparable to that of the
handful of galaxies that have produced more than one Type~Ia supernova. Since
this cluster contains two PAGB A stars, all questions of cluster distance and
reddening drop out, and the fact that their $V$ magnitudes agree to within
0.09~mag decisively argues for a very small range in absolute magnitudes.
Unfortunately, the zero point for this cluster is based at present upon the
old photographic photometry of Harris et~al. (1976).  To remedy this
situation, L. Fullton, STScI summer students S. O'Brien and C. Marois, and
myself are reducing CCD frames obtained with the CTIO 0.9- and 1.5-m
telescopes.  A preliminary color-magnitude diagram is shown in Fig.~3, with
the two PAGB stars shown as large filled circles.  They lie fully 4~mag above
the horizontal branch of the cluster, and are at least a magnitude brighter
than the (rather ill-defined) red-giant tip.  Again, we see vividly that PAGB
A and F stars are the brightest members of \popii{}. 

\begin{figure}
\plotfiddle{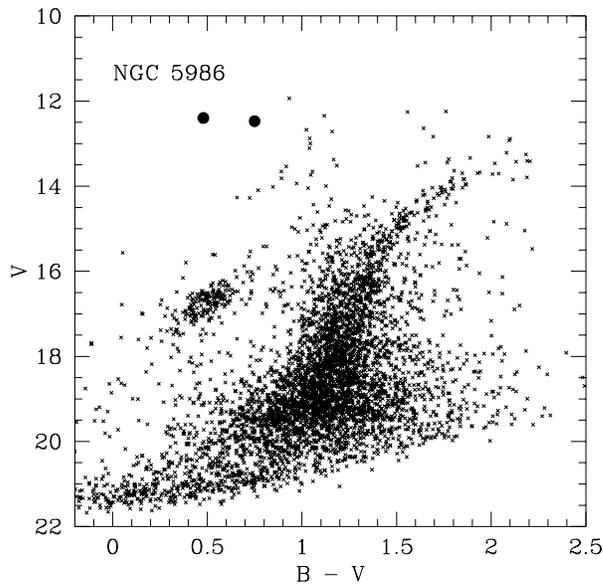}{2.85in}{0.}{40}{40}{-120}{-64}
\caption{Preliminary CCD CMD for the Galactic globular cluster NGC~5986 (Bond, 
Fullton, Marois, \& O'Brien, in preparation). The large filled circles 
mark the two PAGB supergiants in this cluster.  Their $V$ magnitudes agree 
within 0.09~mag.}
\end{figure}

The route to the zero-point calibration for PAGB A and F stars is thus to
begin with subdwarf parallaxes (which the {\it Hipparcos\/} data will soon be
providing) in order to set the distance scale for Galactic globular clusters 
through main-sequence fitting.
This scale will then set the absolute magnitudes for the PAGB stars. Of
course, four PAGB candidates are an insufficient number for such a
calibration, and it will be necessary to find more of them in globular
clusters.  Paradoxically, their very high brightnesses may have hindered their
recognition in the past, since cluster investigators would typically regard
them as foreground stars.  However, as noted below, they can be recognized
very readily using multicolor photometry, and they are so bright that 1-m
class telescopes are more than adequate to survey all of the Galactic globular
clusters. We have now begun such a survey and hope to finish it in 1997.  As
shown in the Appendix below, we expect to find about a dozen PAGB A and F
stars in the Milky Way globular-cluster system, a number roughly comparable to
the number of Galactic Cepheids known in open clusters. 

\section{Discovery Techniques}

Aside from an expected narrow LF, PAGB stars of spectral types A and F have
another advantage: they are very easily recognized.  Since they are stars of
low mass ($\squig0.55M\subsun$) but of high luminosity, they have extremely
low surface gravities (cf.~Table~1).  Around types A and F, the Balmer
discontinuity is very sensitive to $\log g$, and thus the PAGB stars will have
conspicuously large Balmer jumps. 

This is illustrated by Fig.~4, showing spectra I obtained in the 1970's of two
of the field PAGB stars listed in Table~1: \BD\ and HD~46703, whose ultra-low
$\log g$ values were first recognized by Kodaira et~al.\ (1970) and Bond
(1970), respectively. $F_\lambda$~drops by about 1.2~mag across the Balmer 
jump in HD~46703, and fully 1.8~mag in the hotter \BD.

\begin{figure}
\plotfiddle{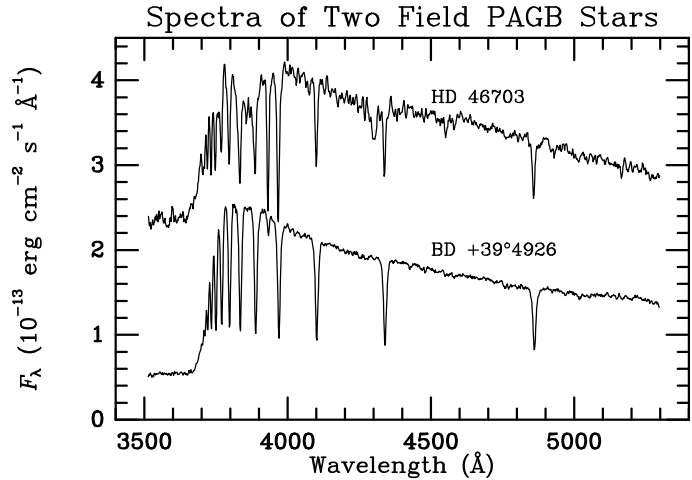}{2.5in}{-90.}{50}{50}{-205}{245}
\caption{Spectra of two field PAGB stars, obtained with the Kitt Peak 
2.1-m telescope and IIDS spectrograph. HD~46703 has been shifted upwards by 
$1.2\times10^{-13}$ for clarity. Note the enormous Balmer jumps of these two 
very low-gravity stars.}
\end{figure}

The large Balmer jumps mean that stars of this type can be recognized easily 
through photometric techniques, if a filter whose bandpass is below the Balmer 
discontinuity is included.  
The classical Str\"omgren {\it uvby\/} system is tailored for 
work of this sort, since its $u$ filter lies entirely below the Balmer jump. 
Examples of the use of this system to detect PAGB stars are the Bond (1970) 
program which, as mentioned, revealed HD~46703, and Bond \& Philip (1973), 
which revealed another field PAGB star, HD~107369. The distinguishing 
characteristic is a high value of the Str\"omgren $c_1$ index.

The Str\"omgren system does, however, have the drawback of low throughput due 
to its narrow bandpasses. In our current work, we are developing a hybrid 
photometric system that combines the Gunn-Thuan $u$ filter with the standard 
Johnson-Kron-Cousins {\it BVI\/} bandpasses.  Gunn $u$ does transmit slightly 
above the Balmer jump, but simulations show that it can nevertheless measure 
the Balmer jump in less telescope time than Str\"omgren $u$, due to its 
greater filter throughput.  Gunn $u$ also has a small red leak, unlike 
Str\"omgren $u$, but this is of little consequence as long as blue stars are 
being measured (and can in any case be subtracted if necessary on the basis of 
$I$ measurements).  

Fig.~5 (kindly prepared by R. Saffer, who is computing theoretical {\it 
uBVI\/} colors using model atmospheres) shows the {\it
uBVI\/} bandpasses along with a theoretical energy distribution for a 9,000~K 
PAGB star.

\begin{figure}
\plotfiddle{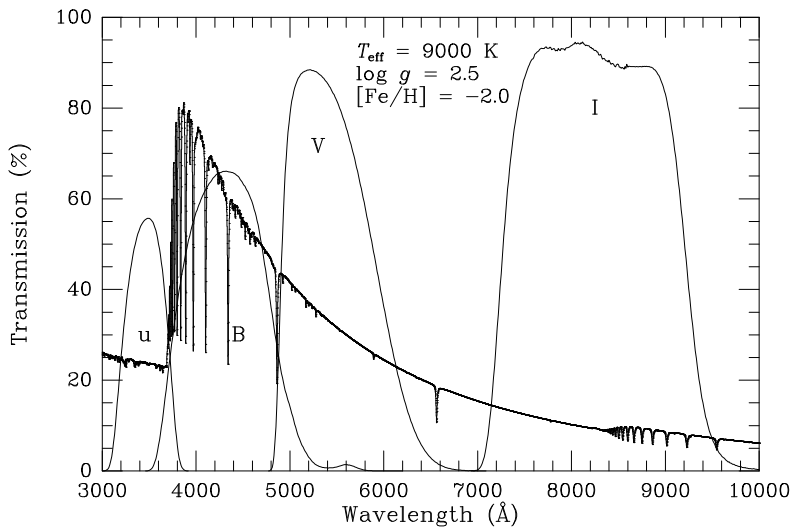}{2.7in}{-90.}{45}{45}{-190}{245}
\caption{Bandpasses for the hybrid {\it uBVI\/} photometric system, superposed
on a theoretical spectrum for a 9,000~K A-type star. The Gunn $u$ filter is
used to detect the extremely large Balmer jumps of PAGB stars at spectral
types of A and F.} 
\end{figure}

\section{Potential Problems}

So far I have painted a rosy picture of a promising new standard candle which 
combines a narrow LF with very easy detectability.  Several potential problems 
must, however, be considered:

\begin{enumerate}

\item {\bf Rarity}. PAGB stars are extremely rare objects. A crude estimate,
based on the time it takes for a star on the $0.546M\subsun$ \Schonberner\
(1983) track to move from $\Teff = 5,000$ to 10,000~K, is that there should be
one PAGB star for every 2,000 red giants.  Here we have defined ``red giant''
as first-ascent stars that lie 1~mag or more above the horizontal branch,
whose lifetime is about $4\times10^7$~yr (Sweigart \& Gross 1978,
0.9~$M\subsun$ track).  Fortunately, with modern CCD techniques, it is readily 
possible to do photometry on tens of thousands of stars in order to find the
rare needles in the haystack. 

\item {\bf Metallicity dependence}. The dependence of PAGB luminosity upon the 
stellar metallicity is set by details of mass loss that are still not well 
understood. Nevertheless, this dependence may be small. Dopita et~al.\ (1992)
give interpolation formulae based on the Vassiliadis \& Wood (1994) 
theoretical PAGB tracks.  This work predicts (with some extrapolation) that at 
a fixed age of 12~Gyr, going from $\log(Z/Z\subsun)=-1$ to $-2$
brightens $M_{\rm bol}$ by 0.3~mag. At a fixed
$\log(Z/Z\subsun)=-1$, reducing the age
from 12 to 8~Gyr brightens the PAGB remnants by 0.1~mag. Thus the effects of 
metallicity (and also age, as long as the population is still reasonably old)
appear to be relatively small, and may even be calibratable.

\item {\bf Variability}. These stars are close to the pulsational
instability strip, and 
indeed there are well-known classes of PAGB variable stars, including 
``UU~Her'' or ``89~Her'' variables (e.g., Bond, Carney, \& Grauer 1984; Fernie 
\& Seager 1995; and 
references therein) as well as the cooler RV~Tauri pulsators. However, by 
staying at spectral types A and F, we should be avoiding most variable stars. 
Indeed, I have monitored the PAGB stars in \wCen\ and NGC~5986 during 
observing runs covering many years, and have never detected variability of 
more than a few hundredths of a magnitude.

\item {\bf Circumstellar extinction}. A significant fraction of the PAGB stars 
in the solar neighborhood were recognized because of their infrared excesses, 
e.g., in the {\it IRAS\/} survey. Thus they are susceptible to circumstellar 
extinction, which could smear out the LF to a large extent. We have, for 
example, obtained a spectacular \HST\/ image of HD~44179 (listed in Table~1, 
and often called the ``Red Rectangle''), showing that the star is in fact not 
seen directly at all.  It lies within a thick dusty disk, and is only seen 
through scattered light.

Fortunately, if we confine ourselves to PAGB stars well out in galactic halos, 
dust formation and extinction may be much less likely. This is for several 
reasons: (1)~dust formation may be difficult in the first place at low 
metallicity; (2)~moreover, the long transition times for low-mass remnants 
give plenty of time for any dust (and gas) that is formed during the AGB phase
to dissipate; (3)~if the AGB ``superwind'' depends on Mira-type pulsation as a
driving mechanism (cf.~Bowen \& Willson 1991), the superwind may not even
occur in very metal-deficient populations that do not produce Miras. 

The small scatter in $M_V$ among globular-cluster PAGB stars (see above) is in
agreement with the above expectation, and indeed field PAGB stars that 
have halo kinematics (e.g., \BD\ and HD~46703) likewise
show little or no evidence for 
surrounding dust.  In fact, it may be that globular-cluster stars also
have a 
difficult time producing PNe.  A survey of all of the Galactic globular 
clusters for PNe revealed only two new ones (G. Jacoby \& L. Fullton, in 
preparation), bringing the total known to only four.  Since the number of PAGB 
stars known in 
globular clusters is already larger than four (if we include the 
four A-F objects listed above
as well as hotter ones such as those tabulated by de~Boer 
1987), in spite of lifetimes comparable to those of PNe, this suggests that 
the typical globular-cluster star 
does not produce a PN during its final evolution. 
Perhaps the few PNe that are seen in globulars derive from blue stragglers, or 
other binary-star interactions.

\item {\bf Alternative evolutionary scenarios}. We have thus far discussed 
hydrogen-burning PAGB stars that are leaving the AGB for the first time.
There are, however, other 
evolutionary paths that could populate this region of the HR diagram in 
galactic halos. A non-exhaustive list would include helium-burning PAGB stars, 
``born-again'' stars that are returning from the top of the white-dwarf 
sequence, and stripped cores in binary systems. Other possibilities include
runaway \popi\ stars (but these would not, at $M_V\approx-3.4$, have the very 
low $\log g$'s of low-mass PAGB stars), and various low-mass stars that fail 
to achieve the AGB due to low envelope masses (e.g.,
the so-called ``AGB-manqu\'e'' and ``PEAGB'' 
stars, but their evolutionary tracks do not attain the high luminosities of 
the genuine PAGB stars and thus may not be a source of confusion).

\item {\bf Poorly understood physics}. Lest the reader think that the 
evolution of PAGB stars is well understood, I mention the extraordinary 
chemical abundances seen in a subset of them, including near-solar C, N, O, S,
and Zn, but strongly depleted iron-group elements (see Bond 1992; Van Winckel
et~al.\ 1995; and references therein). Apparently the material now in the
photosphere at some time in the past
reached a sufficient distance from the star to form grains of
iron-group elements, and then the depleted gas fell back onto the star. 
How this could happen is still a matter of speculation. A possibly related 
phenomenon is that many of the field PAGB stars appear to be binaries, 
typically with periods of about 400--700~days (cf.~Van Winckel et~al.\ 1995).

\end{enumerate}

All of the above concerns emphasize that {\it an empirical calibration of the
PAGB luminosity function and extensive testing in nearby galaxies will be
necessary\/} before they can be applied to the distance-scale problem. 

\section{A Preliminary Test in the Halo of M31}

We have applied our {\it uBVI\/} search technique in the halo of M31, using 
the Kitt Peak Mayall 4-m telescope and a $2048\times2048$ CCD which yields a 
$16'\times16'$ field of view.  We observed three fields at about $40'$--$50'$ 
from the nucleus along the minor axis, in which the surface density of M31
halo red giants is about 10,000 per CCD field. Thus we expect to find about 5
PAGB stars in each field. 

Fig.~6 (left side), kindly supplied by L. Fullton, shows how we select PAGB
candidates. We have plotted a $c_1$-like index, $(u-B)-(B-V)$, vs.\ $B-V$, and
the dashed box isolates the A and F stars with large Balmer jumps (i.e.,
having a large $(u-B)-(B-V)$ index with $0 < B-V < 0.5$). In this field we
find 6 candidates, in almost perfect agreement with expectation. 

\begin{figure}
\plottwo{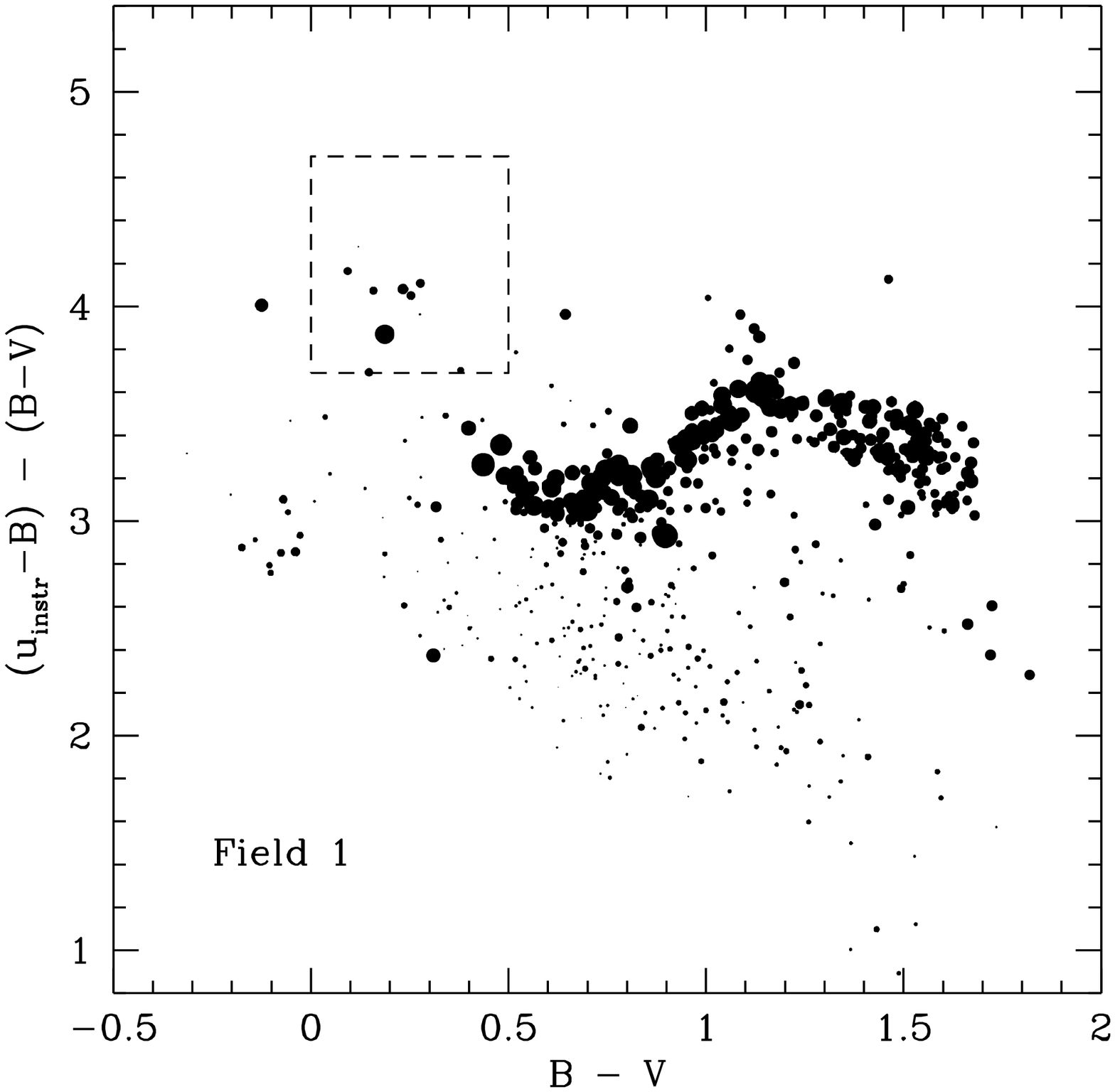}{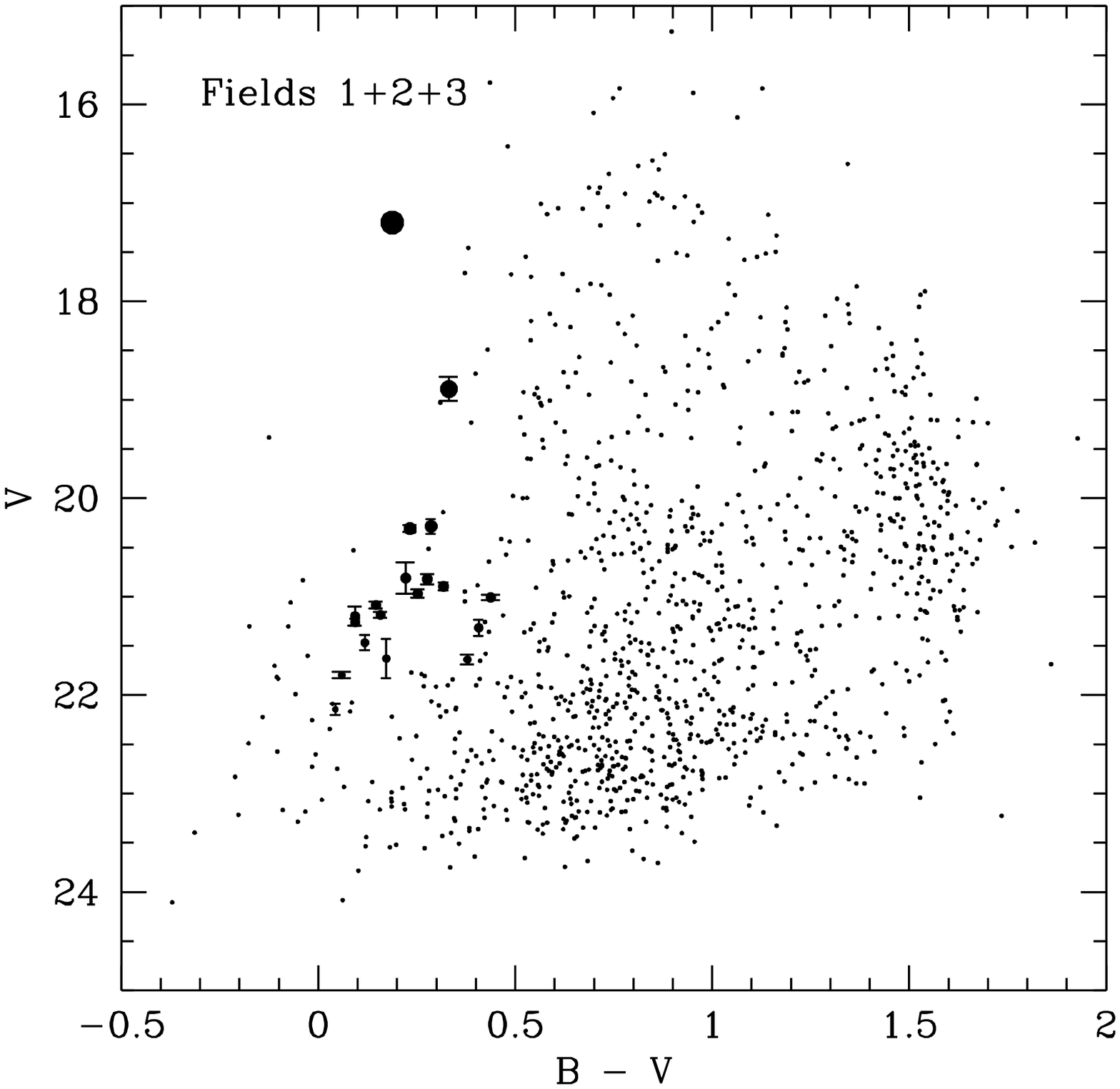}
\caption{{\bf(Left)} A Balmer-jump index, $(u-B)-(B-V)$, vs.\ $B-V$ for a
single $16'\times16'$ field 7~kpc from the center of M31 along the minor axis.
The box isolates PAGB candidates with $B-V$ between 0 and 0.5 and large Balmer
jumps. Dot sizes are proportional to brightness. Only stars detected in all 3
colors are plotted. {\bf(Right)} $V$ vs.\ $B-V$ for PAGB candidates (large
symbols with error bars) in all 3 M31 fields. The candidates (aside from
two bright ones, presumably Galactic foreground horizontal-branch stars) have a
relatively narrow $V$ luminosity function ($\sigma(V)\simeq0.3$~mag) and lie
very close to the $V$ magnitudes expected based on PAGB stars in Milky Way
globular clusters.} 
\end{figure}

Fig.~6 (right side) shows the $V,B-V$ diagram for all three fields, with the
PAGB candidates (those lying inside the dashed box in the left-hand plot)
marked with error bars. Although the reductions are still preliminary at this
writing, these candidates have a mean $V$ of about 21.1 (if we neglect the two
anomalously bright stars, which are presumably foreground horizontal-branch
stars in our own Galaxy's halo).  If we adopt the Galactic calibration of
$M_V=-3.4$, and a foreground reddening of $E(B-V)=0.08$, we find a true
distance modulus $(m-M)_0=24.2$.  This is in superb agreement with the M31
distance modulus of $24.3\pm0.1$ adopted in the review article by van~den
Bergh (1992), and gives us confidence that our method has promise.  It is,
however, troubling that the scatter in the $V$ magnitude is larger than we
expected, but we still need to calibrate the $(u-B)-(B-V)$ index using model
atmospheres. At this stage, some of the objects in the figure may be
background galaxies or QSOs with large Balmer or Lyman jumps, many of which
could be weeded out through a $\log g$ calibration. 

\section{Future Plans}

Much work remains to be done in order to develop these potential excellent 
candles:

\begin{enumerate}

\item {\bf Primary calibration}. As mentioned above, we have begun a {\it
uBVI\/} survey of all of the Galactic globular clusters for PAGB stars. Once
the subdwarf calibration becomes available from {\it Hipparcos\/} parallaxes,
we will have a firm calibration of the PAGB absolute magnitudes.  The PAGB
stars in Milky Way globulars will typically have $V=11$--15, and we expect to
find about a dozen of them (see Appendix). 

\item {\bf Magellanic Clouds}. We have selected PAGB candidates in the 
Magellanic Clouds from Curtis Schmidt objective-prism material obtained at 
Cerro Tololo, and are now reducing {\it uBVI\/} CCD photometry obtained at the 
CTIO 1.5-m telescope. These objects, expected to lie near $V=15.3$, will 
further test and strengthen the $M_V$ calibration.

\item {\bf Local Group}. Moving further out, we will be able to survey fields 
in all of the Local Group galaxies, including both dwarf ellipticals (NGC~147,
NGC~185, NGC~205) and galaxies with Cepheids (M31, M33, NGC~6822, etc.),
allowing a direct confrontation with the Cepheid distance scale. The PAGB
stars in these galaxies will typically have $V\approx21$, as we found in M31
already. 

\item {\bf Intermediate distances}. If the above work is successfully
completed, we would be in a position to move out to intermediate-distance 
galaxies such as those of the Sculptor and M81 groups. At $V=23$--25, the PAGB
stars should be reachable in about one night per field with a 4-m class
telescope.  (Most of the observing time must be spent in the $u$ filter, due
to its relatively low throughput and the low stellar flux below the Balmer
jump.)  We need hardly mention that, for example, the reliable detections of
Cepheids in M81 required a long series of observations with \HST\ (Hughes
et~al.\ 1994). By contrast, the PAGB stars require only one epoch of
observation, and we can go out into the halos where crowding and internal
extinction will be much less severe. 

\item {\bf Virgo Cluster}. Our ultimate aim will be to attain the distance of
the Virgo Cluster, where the PAGB stars should have $V\approx27.5$.  This may
be achievable with \HST\/ and its Advanced Camera, though the signal-to-noise
will be low in a $u$-like filter. 

\end{enumerate}

\section{Summary}

\begin{enumerate}

\item Post-AGB A-F stars appear on theoretical and empirical grounds to be
excellent candidates for standard candles.

\item From a few PAGB A-F stars known to lie in Galactic globular clusters,
they appear to have a small dispersion around $M_V=-3.4$, i.e., 4~mag
brighter than RR~Lyrae stars. {\bf They are the visually brightest members of
Population~II.} 

\item They are very easily recognized via {\it uBVI\/} photometry, due to 
their extraordinarily large Balmer jumps.

\item Their absolute magnitudes can be calibrated in the Milky Way, using
subdwarf parallaxes (from {\it Hipparcos}) to determine distances of globular
clusters that contain PAGB stars, through main-sequence fitting. We have begun
a {\it uBVI\/} survey to find more PAGB stars in Galactic globulars. 

\item Among other advantages, PAGB stars can be observed in spiral halos {\it
and\/} in ellipticals; a time series of observations is not needed; there will
be few problems with crowding or internal reddening; and, due to their narrow
LF, we just need to {\it detect\/} them, rather than doing the
``edge-finding'' necessary for planetary nebulae or the red-giant tip. 

\item We therefore argue that PAGB stars may be {\bf the best available 
Population~II candles}. It should be possible to use them to reach the Virgo 
Cluster in just three steps: subdwarf parallaxes, Galactic globular clusters, 
and then directly to Virgo with \HST.

\end{enumerate}

\begin{acknowledgments}

I gratefully acknowledge support by the NASA UV, Visible, and Gravitational
Astrophysics Program, grant NAGW-4361.  I thank Kitt Peak National and Cerro
Tololo Interamerican Observatories for observing time, and my colleagues for 
their many discussions and hard work.

\end{acknowledgments}

\appendix

\section{The Expected Number of PAGB Stars in the Galactic Globular-Cluster 
System}

We may estimate how many PAGB A-F supergiants should exist in all of the Milky
Way globular clusters using the ``fuel-consumption'' theorem.  For a wide
range of underlying properties, the rate at which stars leave the AGB in a
population whose total luminosity is $L$ is 
$$2\times10^{-11} (L/L_\odot) \, {\rm stars\,yr^{-1}}$$
(see Renzini \& Buzzoni 1986; Ciardullo 1995).

The total luminosity of all Galactic globular clusters (calculated from
the data in Webbink 1985) is
$$2.4\times10^7 L_\odot.$$

Hence the expected number of PAGB A-F stars in all Milky Way globulars is
$$N_{\rm PAGB} \simeq 10 \,\, (\tau_{\rm PAGB}/20,000 \, \rm yr) \,\, stars,$$
where $\tau_{\rm PAGB}$ is the time the PAGB star spends evolving from $T_{\rm
eff}$ = 5,000 to 10,000~K\null. For the $0.546M_\odot$ PAGB track of
Sch\"onberner (1983), $\tau_{\rm PAGB} = 20,000$~yr.

We thus expect to find almost a dozen PAGB stars in the survey that we have 
initiated. The presence of {\it two\/} PAGB stars in the sparse cluster 
NGC~5986 (see above) is a somewhat encouraging, if disquieting, indication 
that our prediction may be an underestimate.  Quite aside from their 
applicability to the extragalactic distance scale, these PAGB stars may 
provide us with surprising new information on the late stages of stellar 
evolution.


\begin{thebibliography}{} 

\bibitem[]{} Aaronson, M., \& Mould, J. 1986, ApJ, 303, 1
\bibitem[]{} Ajhar, E. A., \& Tonry, J. L. 1994, ApJ, 429, 557
\bibitem[]{} Bl\"ocker, T. 1995, A\&A, 299, 755
\bibitem[]{} Bl\"ocker, T., \& Sch\"onberner, D. 1990, A\&A, 240, 11
\bibitem[]{} Bond, H. E. 1970, ApJS, 22, 117
\bibitem[]{} Bond, H. E. 1977, BAAS, 9, 601
\bibitem[]{} Bond, H. E. 1992, Nature, 356, 474
\bibitem[]{} Bond, H. E., Carney, B. W., \& Grauer, A. D. 1984, PASP, 96, 176
\bibitem[]{} Bond, H. E., \& Philip, A. G. D. 1973, PASP, 85, 332
\bibitem[]{} Bowen, G. H., \& Willson, L. A. 1991, ApJ, 375, 53
\bibitem[]{} Brewer, J. P., Richer, H. B., \& Crabtree, D. R. 1995, AJ, 109, 
  2480
\bibitem[]{} Cannon, R.D., \& Stobie, R.S. 1973, MNRAS, 162, 207
\bibitem[]{} Ciardullo, R. 1995, in I.A.U. Highlights of Astronomy \# 10, 
  ed.~I. Appenzeller (Dordrecht, Kluwer), p.~507
\bibitem[]{} de Boer, K. S. 1987, in IAU Colloq.~No.~95, The Second Conference 
  on Faint Blue Stars, ed. A.G.D. Philip et al. (Schenectady, L. Davis Press), 
  p.~95
\bibitem[]{} Dixon, W.V.D., Davidsen, A. F., \& Ferguson, H. C. 1994, AJ, 107, 
  1388
\bibitem[]{} Dopita, M. A., Jacoby, G. H., \& Vassiliadis, E. 1992, ApJ, 389, 
  27
\bibitem[]{} Feast, M. W. 1996, MNRAS, 278, 11
\bibitem[]{} Fernie, J. D., \& Seager, S. 1995, PASP, 107, 853
\bibitem[]{} {Gonzalez}, G., \& {Wallerstein}, G. 1992, MNRAS, 254, 343
\bibitem[]{} Harris, H. C., Nemec, J. M., \& Hesser, J. E. 1983, PASP, 95, 256
\bibitem[]{} Harris, W.E., Racine, R., \& de Roux, J. 1976, ApJS, 31, 13
\bibitem[]{} Hughes, S.M.G. et al. 1994, ApJ, 428, 143
\bibitem[]{} Jacoby, G. H. et al. 1992, PASP, 104, 599
\bibitem[]{} Klochkova, V. G. 1995, MNRAS, 272, 710
\bibitem[]{} Kodaira, K., Greenstein, J.L., \& Oke, J.B. 1970, ApJ, 159, 485
\bibitem[]{} {Lambert}, D. L., {Hinkle}, K. H., \& {Luck}, R. E. 1988, ApJ, 
  333, 917
\bibitem[]{} Lee, M. G., Freedman, W. L., \& Madore, B. F. 1993, ApJ, 417, 553
\bibitem[]{} Luck, R. E., \& Bond, H. E. 1984, ApJ, 279,729
\bibitem[]{} Luck, R. E., \& Bond, H. E. 1989, ApJ, 342, 476
\bibitem[]{} Luck, R. E., Bond, H. E., \& Lambert, D. L. 1990, ApJ, 357, 188
\bibitem[]{} Luck, R. E., Lambert, D. L., \& Bond, H. E. 1983, PASP, 95, 413
\bibitem[]{} Madore, B. F., \& Freedman, W. L. 1995, AJ, 109, 1645
\bibitem[]{} Paczynski, B. 1971, Acta Astr., 21, 417
\bibitem[]{} {Parthasarathy}, M., {Garcia Lario}, P., \& {Pottasch}, S. R. 
  1992, A\&A, 264, 159
\bibitem[]{} Pritchet, C. J. et al. 1987, ApJ, 323, 79
\bibitem[]{} Renzini, A., \& Buzzoni, A. 1986, in Spectral Evolution of 
  Galaxies, ed. G. Chiosi \& A. Renzini (Dordrecht, Kluwer), p. 195
\bibitem[]{} Saha, A., Freedman, W. L., Hoessel, J. G., \& Mossman, A. E. 
  1992, AJ, 104, 1072
\bibitem[]{} Sargent, W. L. W. 1965, Observatory, 85, 116
\bibitem[]{} Sch\"onberner, D. 1983, ApJ, 272, 708
\bibitem[]{} Soria, R. et al. 1996, ApJ, 465, 79
\bibitem[]{} Stanghellini, L., \& Renzini, A. 1993, in IAU Symposium 155, 
  Planetary Nebulae, ed. A. Acker \& R. Weinberger, p.~473
\bibitem[]{} Sweigart, A., \& Gross, P. 1978, ApJS, 36, 405
\bibitem[]{} Tonry, J. L., \& Schneider, D. P. 1988, AJ, 96,807
\bibitem[]{} van den Bergh, S. 1992, PASP, 104, 861
\bibitem[]{} Van Winckel, H., Oudmaijer, R. D., \& Trams, N. R. 1996b, A\&A, 
  312, 553
\bibitem[]{} Van Winckel, H., Waelkens, C., \& Waters, L.B.F.M. 1995, A\&A, 
  293, 25
\bibitem[]{} Van Winckel, H., Waelkens, C., \& Waters, L.B.F.M. 1996a, A\&A, 
  306, L37
\bibitem[]{} Vassiliadis, E., \& Wood, P. R. 1994, ApJS, 92, 125
\bibitem[]{} {Waelkens}, C., {Van Winckel}, H., {Bogaert}, E., \& {Trams}, 
  N. R. 1991, A\&A, 251, 495
\bibitem[]{} {Waelkens}, C., Van Winckel, H., {Trams}, N. R., \& {Waters}, 
  L. B. F. M. 1992, A\&A, 256, L15
\bibitem[]{} Webbink, R.F. 1985, in IAU Symposium 113, Dynamics of Star
  Clusters, ed. J. Goodman \& P. Hut, p.~541 
\bibitem[]{} {Zacs}, L., {Klochkova}, V. G., \& {Panchuk}, V. E. 1995, MNRAS, 
  275, 764

\end{thebibliography}
\end{document}